\documentstyle[psfig]{mn}

\title[The correlation between black hole mass  
and bulge velocity dispersion]{The correlation between black hole mass  
and bulge velocity dispersion in  hierarchical 
galaxy formation models}
\author[Haehnelt\&Kauffmann]{Martin G. Haehnelt and Guinevere Kauffmann\\
Max-Planck-Institut f\"ur Astrophysik, Postfach 1317, 85741 Garching,Germany}
\begin{document}
\maketitle

\begin{abstract}
Recent work has demonstrated that there is a tight correlation 
between the mass of a black hole and the velocity dispersion  
of the bulge of its host galaxy.
We show that the model of Kauffmann \& Haehnelt, in which bulges and   
supermassive black holes both form during major mergers,             
produces a correlation between  $M_{\rm bh}$ and $\sigma$ with slope 
and scatter comparable to the observed relation. 
In the model, the $M_{\rm bh} - \sigma$ relation                                      
is significantly tighter than the correlation between black hole
mass and bulge luminosity or the correlation between bulge
luminosity and velocity dispersion.  
There are two reasons for this: 
i) the gas masses of bulge progenitors depend on 
the velocity dispersion but not on the formation epoch of the bulge, 
whereas the stellar masses of the progenitors depend on both;
ii) mergers between galaxies move black holes along the      
observed $M_{\rm bh } - \sigma$ relation, even at late times when
the galaxies are gas-poor and black holes grow mainly 
by merging of pre-existing black holes. We conclude that 
the small scatter in the observed $M_{\rm bh} - \sigma$ relation 
is consistent with a picture in which bulges and black holes form 
over a wide range in redshift.

\end{abstract}


\section{Introduction}

Several recent papers have shown that a surprisingly tight correlation exists
between the masses of supermassive black holes and the velocity dispersions
of the bulges which host them.
Ferrarese \& Merritt  (2000)  re-analyzed  published samples
of black hole mass estimates, showing that if  the 
analysis is restricted to 12 galaxies with reliable black hole mass measurements,
the correlation is extremely good,                                           
with $M_{\rm bh} \propto \sigma^{5.27\pm(0.4)}$.   
Gebhardt et al. (2000a)  report a similar relation
with a somewhat shallower slope, $M_{\rm bh} \propto \sigma^{3.75(\pm
0.3)}$, and a scatter of only  0.3 dex, based on  a sample of
26 galaxies, including 13 new black hole mass estimates derived using 
Hubble Space Telescope spectra. In a second paper, Gebhardt et
al. (2000b) demonstrate that black holes with reverberation mapping 
mass estimates also fall on the same relation. 

The tightness of this new correlation greatly increases confidence 
in the accuracy of the observed black hole mass estimates. 
It also strengthens theoretical arguments that spheroid formation 
and the growth of black holes are closely linked 
(Richstone et al. 1998; Cattaneo, Haehnelt \& Rees 1999; 
Kauffmann \& Haehnelt 2000 (KH2000); Monaco, Salucci \& Danese 2000; 
Cavaliere \& Vittorini 2000). 
Cattaneo et al. (1999)  and KH2000 
demonstrated that the  $M_{\rm bh } - L_{\rm bulge}$  
relation and the scatter reported by  Kormendy \& Richstone (1995) and
Magorrian et al. (1998) is well reproduced in phenomenological models of galaxy formation  
in hierarchical cosmogonies. In this Letter, we analyze the relation between black hole
mass and bulge velocity dispersion in the KH2000 model.
We assume a $\Lambda$CDM cosmology with 
$\Omega_{\rm mat} = 0.3$,
$\Omega_{_{\Lambda}} = 0.7$, $h= 0.65$, $\sigma_{8} =1$ throughout.

\begin{figure*}
\vspace{-0.5cm}
\centerline{
\hspace{0.0cm}\psfig{file=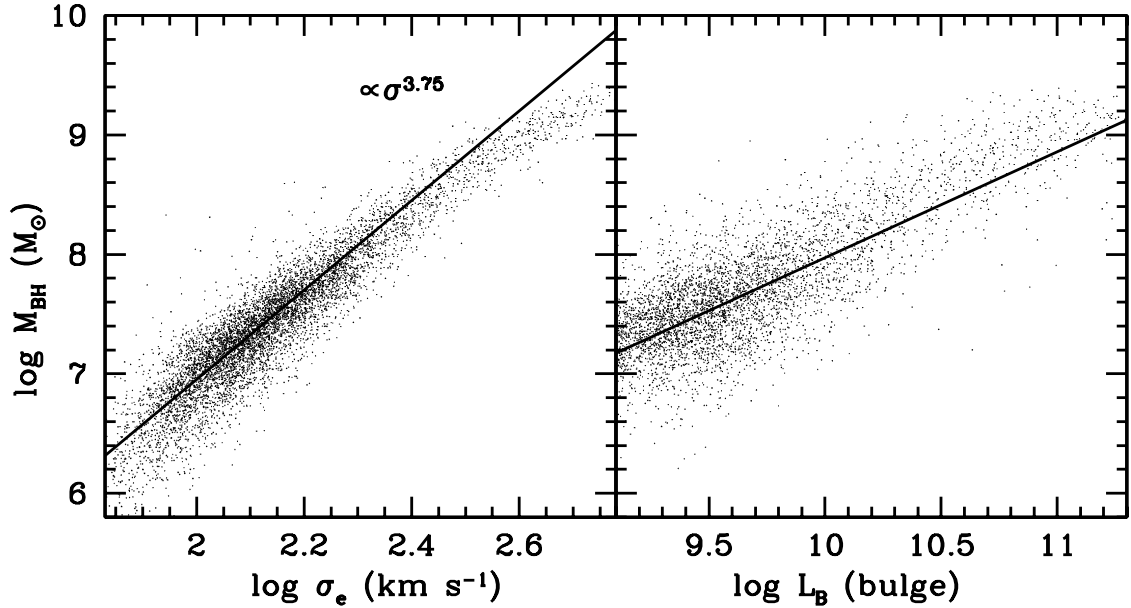,width=16.0cm,angle=0.}
}
\caption{Black hole mass {\it vs} bulge velocity dispersion
(left)  and bulge luminosity (right). Solid lines are the 
fits to an observed sample by Gebhardt et al. (2000).} 
\end{figure*}

\section{The formation and growth of supermassive black holes during
major mergers} 

In cold dark matter (CDM)-like cosmologies, galaxies form in a hierarchy of merging halos
(e.g. White \& Rees 1978; White \& Frenk 1991).
The formation  and evolution of galaxies in such a picture has 
been studied in considerable detail using Monte-Carlo realizations of the 
hierarchical growth of structure, which include simple prescriptions 
to describe gas cooling, star formation and supernova feedback 
(see Kauffmann et al 1999; Somerville \& Primack 1999;
Cole et al 2000 for a selection of recent results). 

Below we summarize the main features of the model described by   
KH2000, who constructed a unified model for the formation
and evolution of galaxies, supermassive black holes and QSOs.
In their model, the quiescent accretion of cooling gas from a halo results
in the formation of a disk. The fraction of cold gas in the disk that forms
stars over one dynamical time is assumed to scale with redshift as $(1+z)^{-3/2}$.
This assumption was required in 
order to fit the observed decline in the total content of cold gas 
in galaxies towards low redshift inferred from observations of damped Lyman-alpha 
systems (Storrie-Lombardi et al. 1996). 
If two galaxies of comparable mass merge, 
a spheroid forms and the remaining gas is transformed into 
stars in a ``starburst''.
The same major mergers are responsible for 
the formation and fuelling of black holes in galactic nuclei. 
During the merger the central
black holes of the progenitors coalesce and a fraction of the available cold gas  
is accreted by the black hole.
The accreted fraction is assumed to scale with the circular velocity $v_{\rm c}$ of the
surrounding dark matter halo as $(1+(280/v_{\rm c})^2)^{-1}$ (This scaling
was adopted in order to fit the slope of the $M_{\rm bh} - L_{\rm bulge}$ 
relation of Magorrian et al (1998).)

The only change in this Letter compared to KH2000 is a 
reduction by a factor of three in the fraction of  
gas assumed to  accrete onto the black hole. This is needed  because the mass    
estimates of black holes from the new data are significantly smaller
than those of Magorrian et al (1998).
In the KH2000 model, black holes grew in mass only during major mergers. 
As pointed out previously,  it is certainly possible that supermassive 
black holes have more complicated accretion histories (see Haehnelt, 
Natarajan \& Rees 1998 and Haehnelt \& Kauffmann 2000 for 
a detailed discussion). Other accretion modes could easily be
incorporated into the model. We note, however, that the 
recent reduction in the estimated total mass density in black holes  
makes arguments for accretion modes
other than those traced by optical- and infrared-bright QSOs less compelling. 

\section{The black hole mass -- bulge velocity dispersion relation} 

Fig. 1a shows scatterplots of black hole mass versus bulge velocity 
dispersion in our models. We do not have a dynamical model for
computing $\sigma$ for the bulges in our simulations.  
For simplicity, we have assumed a constant ratio
$v_{\rm c}/ \sigma$, where $v_{\rm c}$ is the circular velocity of 
the halo in which the bulge found itself after its last major merger. 
It is easy to show that any well-mixed population of stars orbiting within an
isothermal potential must satisfy 
$v_{\rm c}/\sigma = \sqrt{2}$. 
Following Kauffmann \& Charlot (1998),   
we choose the ratio of $v_{\rm c}/\sigma$ to reproduce the zero point of the
observed Faber-Jackson relation (see Fig. 2). To fit the observations,
we require $v_{\rm c}/\sigma \sim 2$.

The thick solid  line in fig. 1a ($M_{\rm bh}\propto \sigma^{3.75}$) 
shows the relation derived by Gebhardt et al. (2000a),  which is reproduced 
very well. In the models, the total mass of gas that cools in a halo 
scales roughly as $\sigma^3$. Feedback effects steepen the relation
to approximately  $M_{\rm bh}\propto \sigma^4$, but feedback would have to be  
more extreme in order to obtain a correlation  as steep as 
$M_{\rm bh} \propto \sigma^{5.3}$ (Ferrarese \& Merritt 2000).  
The scatter in the relation is 0.2 dex, somewhat smaller than the 0.3 dex  
reported by Gebhardt et al. which include measurement errors. 
Fig. 1b shows  
the black hole mass -- bulge 
luminosity correlation compared with the observed relation 
in the Gebhardt et al. sample.  For the                 
simulated galaxies, the scatter is about a factor two larger and again 
agrees well with the observational data. The large scatter in the 
relation in Fig. 1b is a consequence of the large dispersion in the relation between 
the luminosities and  velocity dispersions of bulges. This is illustrated in       
in Fig 2a where we plot the $L_{\rm bulge} -\sigma$ (Faber-Jackson) 
relation for the elliptical 
galaxies in our model.
Elliptical galaxies are defined as those objects with a ratio of bulge
to total luminosity greater than 0.4 in the B-band.  
In the next section, we study the  origin of the scatter in these relations and explain
why the scatter in the $M_{\rm bh} - \sigma$ relations  
is significantly 	 smaller than that of the $L_{\rm bulge} - \sigma$ or 
$ M_{\rm bh} - L_{\rm bulge}$ relations.
\begin{figure*}
\vspace{-0.5cm}
\centerline{
\hspace{0.0cm}\psfig{file=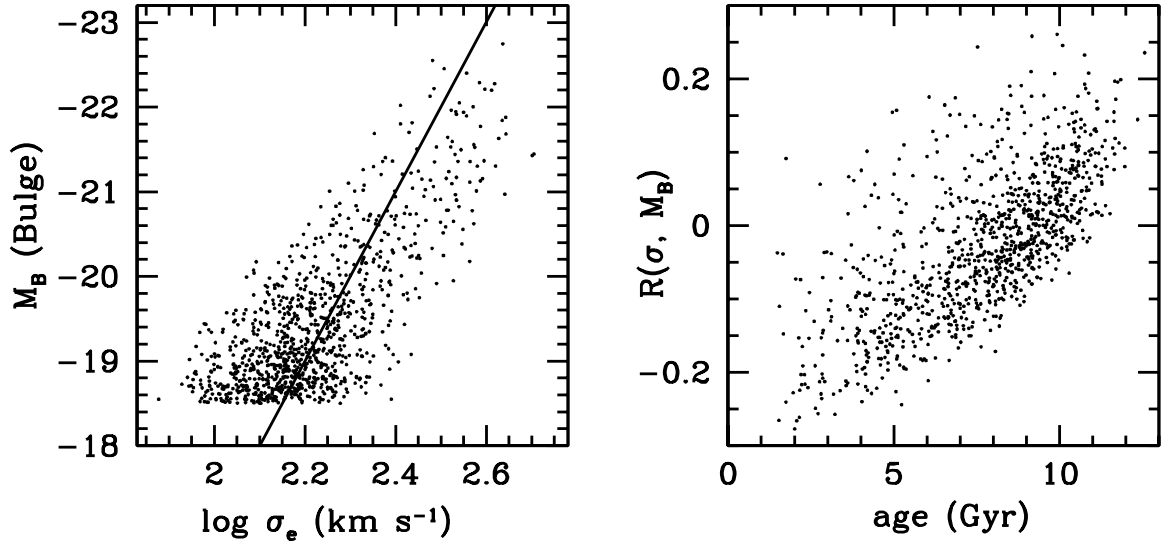,width=16.0cm,angle=0.}
}
\vspace{0.0cm}
\caption{{\it Left:} The Faber-Jackson relation between 
bulge absolute magnitude in the B-band and stellar velocity dispersion. The solid line  
is the relation $\log \sigma= -0.102 M_B +0.24$ obtained by 
Forbes \& Ponman (1999) from their best fit to the elliptical 
galaxies in the sample of 
Prugniel \& Simien (1996) {\it Right:} The residuals 
$R(\sigma, M_B)  = \log{\sigma} + 0.102 M_B -0.24$ in the
Faber-Jackson relation as a function of the age of the bulge.} 
\end{figure*}

\section{The scatter in the $M_{\rm bh}-\sigma$ relation}

Previous work has demonstrated that hierarchical galaxy formation
models  succesfully reproduce the Faber-Jackson ($L_{\rm bulge}
-\sigma$) and the Magorrian ($M_{\rm bh} - L_{\rm bulge}$) 
relations (Kauffmann \& Charlot 1998, KH2000).  The
slopes of these correlations are determined by how much gas is able to cool and
forms stars or is funneled to the centre in a dark matter halo of 
given mass or circular velocity.
This is set by the assumed balance between gas cooling, star formation,
supernova feedback and accretion in the halos.
The correlations are tight for bulges that form at fixed epoch, but
the scatter increases substantially if the galaxy population
as a whole is considered. This is demonstrated by Fig. 2b, 
where we plot the residuals in our ``Faber-Jackson'' relation as 
a function of the age of the bulge. Age is defined as the
time elapsed since the bulge had its last major merger. 
As can be seen, the residuals in the Faber-Jackson relation correlate 
strongly with age: young bulges of fixed velocity dispersion are
brighter than old bulges. Our results are in good agreement with those  
of Forbes \& Ponman (1999), who study residuals in the 
Faber-Jackson relation for a sample of 88 nearby ellipticals with ages
determined from absorption line spectroscopy.
KH2000 showed that a similar effect is expected for the 
$M_{\rm bh} - L_{\rm bulge}$ relation.
Young bulges of fixed luminosity contain less massive black holes 
than older bulges. This prediction was recently confirmed by 
Merrifield, Forbes \& Terlevich (2000).  

Why does the $M_{\rm bh} - \sigma$ relation exhibit such small scatter?        In Fig. 3a, we show how the gas and the  stellar masses of bulge progenitors 
vary as a function of the formation redshift of the bulge.      
Results are shown for bulges with $\sigma = 150$ km s$^{-1}$ and
$\sigma = 200$ km s$^{-1}$.  The cold gas masses 
of the progenitors of  bulges of given velocity dispersion
{\em are essentially independent of redshift}, whereas their stellar masses
decrease at high redshift. As discussed by KH2000, this decrease in 
the ratio of gas mass to stellar mass in galaxies is crucial for explaining  
the observed rapid decline of the QSO activity at late times, as 
well as the decrease in the total content of cold gas in galaxies 
inferred from damped Ly$\alpha$ systems (Storrie-Lombardi et al. 1996).

In our model  the amount of gas available for accretion
onto a black hole during a major merger depends only on 
the velocity dispersion of the bulge and
not on redshift. This together with our assumption 
that a fixed fraction of this gas is accreted by the black hole
accounts for the small scatter in the $M_{\rm bh} - \sigma$ relation 
for bulges forming in gas-rich mergers.

What about gas-poor mergers, where gas accretion contributes little to the
growth of the black hole?  Such gas-poor mergers are important 
for massive bulges forming at late times (KH2000). 
Figure 3b shows ``evolutionary tracks'' in the $M_{\rm bh}- \sigma$ plane for 20 black
holes in our model with masses $\sim 10^9 M_{\odot}$. These 
black holes typically form in     
3-5 merging events spaced quite widely in redshift. At late times,
the host galaxies of these black holes have rather small 
cold gas fractions. Nevertheless, it is clear from our plot that black holes 
always move along the correlation as they merge, 
even for massive systems at late times.
The latter  can be understood as follows. 
Let us consider the case where black holes grow only by merging and
therefore have masses that scale in proportion to those of their host halo.
For  halos of constant characteristic density, the velocity dispersion
$\sigma$ scales $\propto M_{\rm halo}^{1/3}$. With decreasing redshift 
the characteristic density decreases and the velocity dispersion grows 
more slowly with increasing mass: $\sigma \propto \rho^{1/6} 
M_{\rm halo}^{1/3}$. 
In CDM-like cosmogonies the characteristic mass of typical dark matter
haloes scales as $\propto \rho^{-2/(3+n)}$, where $n$ is the effective 
slope of the DM fluctuation spectrum. At galaxy 
scales  $n\sim -2$. This gives a scaling $M_{\rm bh} \propto
\sigma^{4}$.

\begin{figure*}
\vspace{-0.5cm}
\centerline{
\hspace{0.0cm}\psfig{file=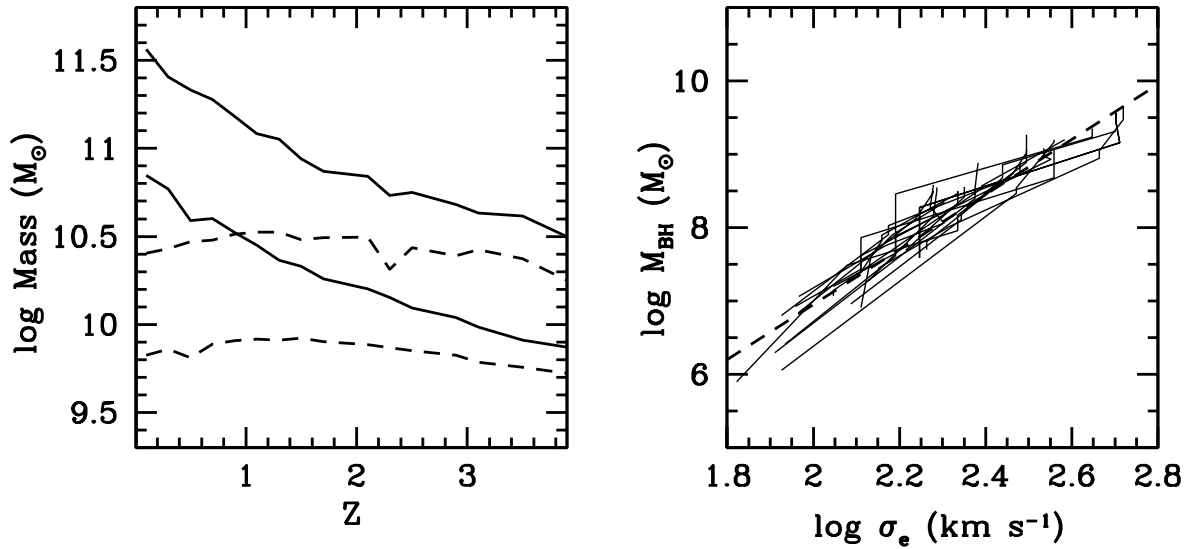,width=16.0cm,angle=0.}
}
\vspace{0.0cm}
\caption{{\it Left:} Gas (dashed curves) and  stellar (solid curves) masses 
of bulge progenitors 
as a function of formation redshift for bulges with 
$\sigma = 150$ km s$^{-1}$ (lower curves)  and 200 km s$^{-1}$ (upper curves).
{\it Right:}  ``Tracks'' of bulges in the black hole mass bulge velocity 
dispersion plane. The dashed line is the observed relation 
from Gebhardt et al. (2000a).}  
\end{figure*}

\section{Discussion and Conclusions} 

We have demonstrated that the model of KH2000       
where supermassive black holes are formed and fuelled
during major mergers
produces a tight correlation between black hole mass 
and bulge velocity dispersion very similar to that observed
by Gebhardt et al. (2000a). The model also reproduces the observed relation
between black hole mass and bulge luminosity, as well as the relation
between  bulge luminosity
and bulge velocity dispersion.   

In the model, the slopes of these relations are determined 
by how much gas cools, forms stars or is funneled to the 
centre in dark matter halos of different mass/velocity dispersion. 
Adopting simple, but physically plausible 
prescriptions to describe these processes, 
we are able to fit all three relations simultaneously. 
Note that the only change we have made compared to KH2000 
is an overall reduction of the accreted gas fraction. 

The large scatter in the $M_{\rm bh} - L_{\rm bulge}$ and the $L_{\rm
bulge} - \sigma$ relations is primarily a consequence of the wide range
in redshift over which  bulges and supermassive black holes form. 
The stellar mass of a bulge of fixed velocity dispersion                  
depends strongly on when the bulge formed. Bulges that form early 
are less massive than bulges that form late. Note that the
same is true of disk galaxies in our model. These predicted correlations
with age appear to be supported by the available data.   
%Vogt et al (1996) have already shown that the B-band Tully-Fisher relation 
%remains constant out to redshift 1, which indicates that the stellar
%masses of disks of fixed circular velocity  must have 
%been smaller in the past.

Nevertheless, the scatter in the $M_{\rm bh}-\sigma$ relation is
small. There are two reasons for this.
\begin{itemize} 
\item{The amount of gas accreted by a black hole during the formation
of a bulge of fixed velocity dispersion does not depend on redshift. 
In our model, disk galaxies of given circular velocity contain about 
the same amount of gas at all redshifts. This is a  further prediction
of the model to be tested by future observations}. 
\item{The frequent  merging of galaxies in hierarchical cosmogonies moves
galaxies along the correlation in the $M_{\rm bh} - \sigma$ plane, even 
when galaxies are gas-poor and their black holes grow mainly 
by the merging of pre-existing black holes}.
\end{itemize}

The model of KH2000 is a supply-limited model for the accretion
history of supermassive black holes. We note that the model does 
not allow for any dependency in the amount of gas accreted by 
the black hole on the orbital parameters
of the merger. We have also assumed a fixed scaling between the 
velocity dispersion of a bulge and the circular velocity of the halo 
in which it resides. These are all potential sources of additional
scatter. Detailed  numerical simulations will be needed to decide 
whether a physical mechanism  that depends on the bulge potential
and that limits the growth of supermassive black holes is required 
to keep the correlation tight. From our present analysis,  we conclude
that the tightness of the observed correlation is in apparent
agreement with a scenario in which merging events produce supermassive
black holes and bulges over a wide range in redshift.

%\acknowledgements

\end{document}